\documentclass[letterpaper, 10 pt, conference]{ieeeconf}  

\IEEEoverridecommandlockouts

\usepackage{cite}
\usepackage{amsmath,amssymb,amsfonts}
\usepackage{algorithmic}
\usepackage{graphicx}
\usepackage{textcomp}
\usepackage{xcolor}
\usepackage{nccmath}

\usepackage{graphics} 
\usepackage{subcaption}
\usepackage{url}

\usepackage{multirow}

\usepackage{lipsum}  
\newtheorem{defi}{Definition}[section]

\newtheorem{lem}[defi]{Lemma}

\title{\LARGE \bf
Introducing Graph Learning over Polytopic Uncertain Graph
}

\author{Masako Kishida$^{1}$ and Shunsuke Ono$^{2}$
\thanks{This work was accepted to be presented at the Graph Signal Processing Workshop 2024}
\thanks{*This work was supported by JST, PRESTO Grant Number JPMJPR22C3, Japan.}
\thanks{$^{1}$Kishida is with Principles of Informatics Research Division,
        National Institute of Informatics, Tokyo, Japan
        {\tt\small kishida@nii.ac.jp}}
\thanks{$^{2}$Ono is with the Department of Computer Science, Tokyo Institute of Technology, 
Tokyo, Japan
        {\tt\small ono@c.titech.ac.jp}}
}

\begin{document}

\maketitle
\thispagestyle{empty}
\pagestyle{empty}

\begin{abstract}

This extended abstract introduces a class of graph learning applicable to cases where the underlying graph has polytopic uncertainty, i.e., the graph is not exactly known, but its parameters or properties vary within a known range. By incorporating this assumption that the graph lies in a polytopic set into two established graph learning frameworks, we find that our approach yields better results with less computation.

\end{abstract}

\section{Introduction}\label{sec:intro}
A class of graph signal processing (GSP) problems \cite{ShuNF13, SanM13, Ort22} focuses on obtaining the underlying graph of a graph signal. This area, known as graph learning, aims to identify the graph topology from the observed data  \cite{DonTF16, Kal16}. It has been studied extensively not only in the signal processing community but also in the machine learning community. 

In this extended abstract, we address the problem of graph learning for graphs that are known to reside within a polytopic set.
Polytopic uncertainty, which generalizes interval uncertainty, has been extensively studied in control theory for its computational and expressive capabilities \cite{BoyEF94, VlaJ14}.
Applying this concept to graph learning allows for the incorporation of prior knowledge about various graph properties, including those with edge connections that depend on uncertain parameters or time-varying structures. 
In particular, we incorporate these constraints into the formulations proposed in \cite{DonTF16} and  \cite{Kal16}.
Our analysis shows that these formulations lead to convex optimizations with a reduced number of parameters, and we demonstrate that this approach leads to improved results.

\subsection*{Notation}
A vector $\mathbf{1}$ represents the vector of ones.
For a vector $v\in \mathbb{R}^n$, $v\geq0$ indicates the all elements are nonnegative.
For a matrix $A$, $\|A\|_F=\sqrt{ \text{Tr}(AA^\top)}$ is the Frobenius norm of $A$, and $\|A\|_{1,1}=\sum_{i,j}|a_{ij}|$ is the element-wise norm-1 of $A$.
For matrices $A$ and $B$ of the same size, $A \circ B$ denotes the Hadamard product.

\section{Polytopic uncertain graph}

We consider a weighted undirected graph $\mathcal{G} = (\mathcal{N}, \mathcal{E}, W)$, where $\mathcal{N}$ represents the set of nodes, $\mathcal{E}$ represents the set of edges and $W$ is the adjacency matrix. 
A graph Laplacian matrix $L$ is defined as $L = \text{diag}(W\mathbf{1})- W$.
Let $n=|\mathcal{N}|$ denote the number of nodes.

A graph Laplacian of the polytopic uncertain graph $\mathcal{G}$, $L(\mathcal{G})$, resides within the convex hull of $p$ given graph Laplacians $L_i$ for $i = 1, \cdots, p$:
\begin{align}\begin{aligned} \label{eq:polyset}
L(\mathcal{G} )& \in \mathbf{L} \\
&:= \{L\in\mathbb{R}^{n \times n}: L = \medop\sum_{i=1}^{p} \theta_i L_i, \ \theta \geq 0, \ \mathbf{1}^\top \theta = 1\}.
\end{aligned}\end{align}
$\mathbf{L}$ is a set of matrices representing the set of all possible graph Laplacians under consideration. 
The graph Laplacian of a polytopic uncertain graph is determined by a specific combination of $L_i$ for $i = 1, \cdots, p$, 
i.e., once  $\theta = \left[\theta_1, \cdots, \theta_p\right]^\top$ is fixed.

\begin{lem}
Any matrix $L\in \mathbf{L}$ is a graph Laplacian if $L_i$, $i = 1, \cdots, p$, are all graph Laplacians.
\end{lem}

Similarly, we can consider the set of weighted adjacency matrices for given $p$ weighted adjacency matrices $W_i$ for $i = 1, \cdots, p$:
\begin{align}\begin{aligned} \label{eq:polyset2}
W(\mathcal{G} )& \in \mathbf{W} \\
&:= \{W\in\mathbb{R}^{n \times n}: W =\medop \sum_{i=1}^{p} \theta_i W_i, \ \theta \geq 0, \ \mathbf{1}^\top\theta = 1\}.
\end{aligned}\end{align}

Let define the set of weight parameters:
\begin{align}
\Theta =\{\theta\in\mathbb{R}^p:  \theta \geq 0, \ \mathbf{1}^\top \theta=1\}.
\end{align} 
\section{Proposed problem formulations}
Here, we consider two problem settings to learn a graph; learning a graph Laplacian \cite{DonTF16}, and learning an adjacency matrix \cite{Kal16}. 
Both approaches assume that the signal is smooth in the sense that $\medop \sum_{i, j} W_{ij}\|x_i-x_j\|^2$, where $x_i$ and $x_j$ are signals at nodes $i$ and $j$, is small.


\subsection{Learning graph Laplacian}\label{sec:dong}
The following problem is considered in \ \cite{DonTF16}, 
\begin{align}\begin{aligned}\label{eq:dong}
&\min_{L}  \text{Tr}(X^\top LX ) +\alpha   \left\| L\right\|_F^2\\
\text{s.t. } &
\text{Tr}(L)= n, \\ 
& L_{ij} =L_{ji} \leq 0, \ i\neq j\\
& L\mathbf{1}= 0,
\end{aligned}\end{align}
with $\alpha> 0$. Here, $X$ is a $n$ by $m$ matrix that contains the $m$ input data samples, i.e., $i$th row represents a signal on the $i$th nodes, and the $j$th column represents the $j$th data sample.

Adding the constraint $L\in \mathbf{L}$, where  $\mathbf{L}$ is defined in \eqref{eq:polyset},  leads to the following problem:
\begin{align}\begin{aligned}\label{eq:dong1}
\min_{\theta \in \Theta} \ & \text{Tr}\left(X^\top \left(\medop\sum_{i=1}^{p} \theta_iL_i \right) X \right) +\alpha   \left\| \medop\sum_{i=1}^{p} \theta_i L_i\right\|_F^2.
\end{aligned}\end{align}
This can be expressed as
\begin{align}\begin{aligned}\label{eq:dong2}
&\min_{\theta \in \Theta} \  \theta^\top u +\alpha  \theta^\top \Phi \theta\\
\text{s.t. } &\theta^\top \ell = n, 
 \end{aligned}\end{align}
where
\begin{align}\begin{aligned}
 u_i &= \text{Tr}\left(X^\top L_i X \right), \ i = 1, \dots, p,\\ 
 \Phi_{ij} &=  \Phi_{ji}=\text{Tr}\left( L_i^\top L_j \right), \ i \geq j  =  1, \dots, p,\\
  \ell_i &= \text{Tr} \left( L_i \right).
\end{aligned}\end{align}
This is a convex quadratic program with only $p$ parameters subject to linear constraints, which can be easily solved. 
It is convex because $ \Phi$ is positive semidefinite by construction. 
\subsection{Learning adjacency matrix}\label{sec:kal}
The following problem is considered in \cite{Kal16}, 
\begin{align}\begin{aligned}\label{eq:kal}
&\min_{W}  \|W\circ Z\|_{1,1}-\alpha   \mathbf{1}^\top \text{log}(W\mathbf{1}) +\frac{\beta}{2}\left\| W\right\|_F^2\\
\text{s.t. } &
W_{ij}\geq0, \ i,j =1,\dots,n,\\
&W = W^\top, \\ 
& \text{diag}(W)=0,
\end{aligned}\end{align}
with $\alpha> 0$ and $\beta>0$. 
Here, $Z$ is the pairwise distances matrix whose $i, j$ element is defined by $Z_{ij} = \|x_i-x_j\|^2$.

Adding the constraint $W\in \mathbf{W}$, where $\mathbf{W}$ is defined in \eqref{eq:polyset2},  leads to the following problem:
\begin{align}\begin{aligned}\label{eq:kal1}
\min_{\theta}  \!& \left\|\medop\sum_{i=1}^{p} \theta_iW_i \! \circ \!Z\right\|_{1,1}\!\!\! \!\!\!\!   - \!  \alpha \mathbf{1}^\top \text{log}\left(\!\medop\sum_{i=1}^{p} \theta_iW_i \mathbf{1}\!\right)
\!+\!\frac{\beta}{2}\!  \left\| \medop\sum_{i=1}^{p} \theta_i W_i\right\|_F^2\!\!.
\end{aligned}\end{align}
This can be expressed as
\begin{align}\begin{aligned}\label{eq:kal2}
\min_{\theta\in \Theta}  \theta^\top v -\alpha   \mathbf{1}^\top \text{log}\left( \theta^\top w \right)+\frac{\beta}{2} \theta^\top \Psi  \theta
\end{aligned}\end{align}
where
\begin{align}\begin{aligned}
 v_i &=\|W_i\circ Z\|_{1,1} , \ i = 1, \dots, p,\\ 
w_i&=  W_i\mathbf{1}, \ i = 1, \dots, p,\\
 \Psi_{ij} &=  \Psi_{ji} =\text{Tr}\left( W_i^\top W_j \right), \ i \geq j  =  1, \dots, p.
\end{aligned}\end{align}
Because the logarithm of a sum is concave, this is again a convex program.

\section{Numerical examples}\label{sec:ex}

\begin{figure}[htbp]
  \centering
\includegraphics[width=1\linewidth, viewport=230 30 2000 540, clip]{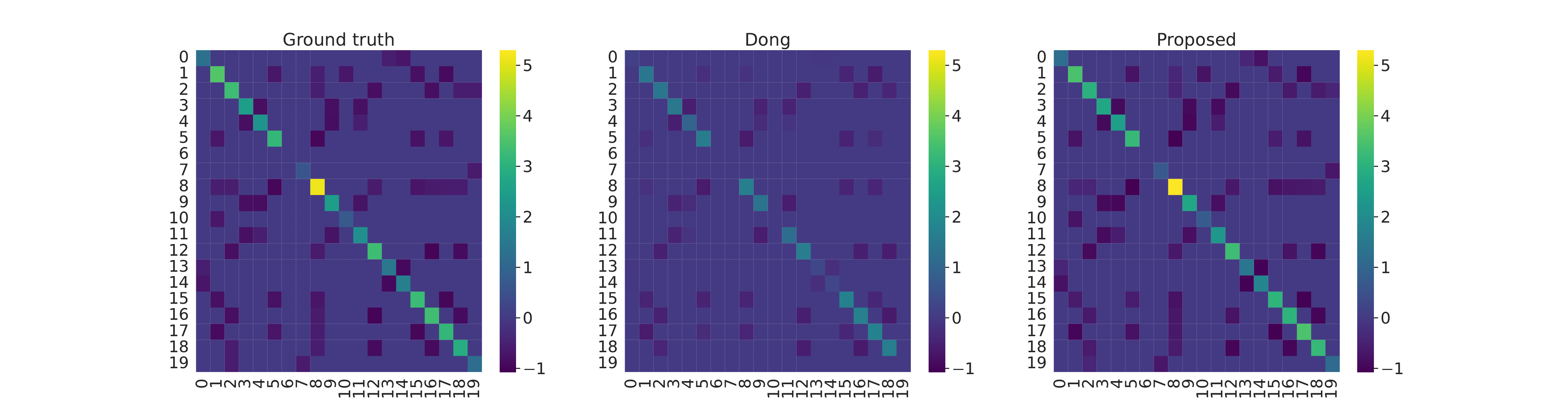} \\
\includegraphics[width=1\linewidth, viewport=230 30 2000 540, clip]{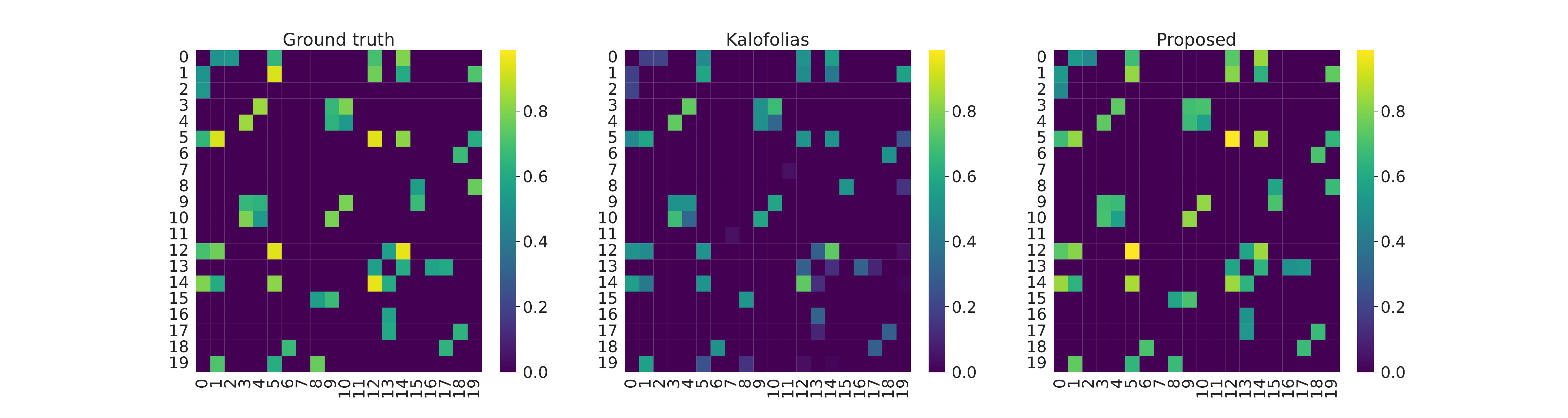} 
  \caption{Matrix heatmaps: 1st row for Section \ref{sec:dong}, 2nd row for Section \ref{sec:kal}, left for ground truth, middle for \cite{DonTF16} and \cite{Kal16}, right for proposed.   }\label{fig:1}
  \vspace{-.2in}
\end{figure}

Following \cite{DonTF16, Kal16}, we simulated the approaches proposed by these authors and our proposed approaches using a random geometric graph with 20 nodes and 100 Tikhonov type signals. Known matrices, $L_1$ and $L_2$ (similarly $W_1$ and $W_2$), were chosen so that a convex combination of them is the ground truth. The results are summarized in Table \ref{table:1} and Figure \ref{fig:1}. We observed that although the proposed approach did not retrieve the ground truth, it performed well, resulting in small Frobenius norms as well as perfect normalized mutual information (NMI) scores and F-measures.

\begin{table}[h!]
\centering
\begin{tabular}{ |p{2cm}||p{1.3cm}|p{1.3cm}|p{1.3cm}|  }
 \hline
 &NMI & F-measure&  $\| \cdot\|_F$\\
 \hline
Dong et. al\cite{DonTF16}&   0.901 &  0.740  &7.701\\
Proposed \eqref{eq:dong2} &1.0&1.0&   1.276\\
 \hline
Kalofolias \cite{Kal16}&   0.931  & 0.787   &2.367\\
Proposed \eqref{eq:kal2} &1.0 &1.0&  0.415\\
 \hline
\end{tabular}
\caption{Errors between the ground truth matrix and learned graph}
\label{table:1}
\end{table}

\section{Conclusions}\label{sec:con}
In this abstract, we discussed how we can integrate polytopic uncertainties in graph learning to reflect prior knowledge of the underlying graph. The proposed formulations allow us to identify the underlying graph in a computationally efficient manner with far fewer parameters to optimize. Future work will investigate real applications as well as combining polytopic uncertain graphs with other graph learning approaches, signal separations, and signal recovery. It is also of interest to investigate the relationship between the smooth signal requirement and a polytopic set.

\bibliographystyle{IEEEtran}
\bibliography{IEEEabrv,myref}

\end{document}